\documentclass[12pt]{iopart}
\usepackage[svgnames,table]{xcolor}
\usepackage[utf8]{inputenc}
\usepackage{appendix}
\usepackage{fouriernc}
\usepackage{float}
\usepackage{soul}
\usepackage{graphicx}
\usepackage{dcolumn}
\usepackage{bm}
\usepackage{caption}
\usepackage{appendix}
\usepackage{parskip}
\usepackage{multirow}
\usepackage{indentfirst}
\usepackage{hyperref}
\hypersetup{
        colorlinks=true,
        citecolor=SteelBlue,
        filecolor=LimeGreen,
        linkcolor=SlateBlue,
        urlcolor=Purple
}
\usepackage[normalem]{ulem}
\expandafter\let\csname equation*\endcsname\relax
\expandafter\let\csname endequation*\endcsname\relax
\usepackage{amsmath}
\usepackage{amssymb}
\usepackage{array}

\definecolor{lightcol}{RGB}{245,245,245}
\begin{document}
\def\no{\nonumber \\}
\def\no{\quad \\}
\def\noQ{\nonumber \\}
\setlength{\parindent}{1em}
\setlength{\parskip}{.5em}


\title[]{Identifying noise transients in gravitational-wave data arising from nonlinear couplings}

\author{Bernard Hall$^{1}$, Sudhagar Suyamprakasam$^{2,3}$, Nairwita Mazumder$^{1}$, Anupreeta More$^{2}$,
Sukanta Bose$^{1,2}$
}
\address{$^1$ Department of Physics \& Astronomy, Washington State University
	1245 Webster, Pullman, WA 99164-2814, U.S.A}
\address{$^2$ Inter-University Centre for Astronomy and Astrophysics, Post
  Bag 4, Ganeshkhind, Pune 411 007, India}
\address{$^3$ Nicolaus Copernicus Astronomical Center, Polish Academy of Sciences, Bartycka 18, 00-716, Warsaw, Poland}

\ead{sukanta@wsu.edu}

\vspace{5pt}
\begin{indented}
\item[]\today
\end{indented}

\begin{abstract}
Noise in various interferometer systems can sometimes couple non-linearly to create excess noise in the gravitational wave (GW) strain data. Third-order statistics, such as  bicoherence and biphase, can identify these couplings and help discriminate those occurrences from astrophysical GW signals. However, the conventional analysis 
can yield large bicoherence values even when no phase-coupling is present, thereby, resulting in false identifications. Introducing artificial phase randomization in computing the bicoherence reduces such occurrences with negligible impact on its effectiveness for detecting true phase-coupled disturbances. We demonstrate this property with simulated disturbances -- focusing only on short-duration ones (lasting up to a few seconds)
and employing mainly the auto-bicoherence in this work. Statistical hypothesis testing is used for distinguishing phase-coupled disturbances from non-phase coupled ones when employing the phase-randomized bicoherence. {We also obtain an expression for the bicoherence value that minimizes the sum of the probabilities of false positives and false negatives. This can be chosen as a threshold for shortlisting bicoherence triggers for further scrutiny for the presence of non-linear coupling.} Finally, the utility of the phase-randomized bicoherence analysis in GW time-series data is demonstrated for the following three scenarios: (1) Finding third-order statistical similarities within categories of noise transients, such as blips and koi fish. If these non-Gaussian noise transients, or glitches, have a common source, their bicoherence maps can have similarities arising from common bifrequencies related to that source. (2) Differentiating linear or non-linear phase-coupled glitches from  compact binary coalescence signals through their bicoherence maps. This is explained with a simulated signal. (3) Identifying repeated bifrequencies
in the second and third  observation runs (i.e., O2 and O3) of LIGO and Virgo. 
\end{abstract}

\maketitle

\section{\label{sec:1_intro}Introduction}

The detectors of the Laser Interferometer Gravitational-wave Observatory (LIGO)~\cite{aLIGO2010}, Virgo~\cite{virgo} and the Kamioka Gravitational Wave Detector (KAGRA)~\cite{kagra} are among the most sophisticated scientific instruments in the world. To date, LIGO and Virgo have detected several dozens of gravitational wave (GW) signals \cite{GWTC-1, GWTC-2, GWTC-2-1, GWTC-3} from colliding binaries involving black holes and neutron stars. These detectors on the globe are  sensitive to GW signals in the $10 - 10^4$ Hz band~\cite{aLIGO2015}, but are often plagued by unwanted noise transients from various terrestrial sources, which can adversely impact that sensitivity~\cite{Davis-2021, Acernese-2023, joshi, sunil}. The quality of data or search sensitivity can be improved by removing or minimizing the noise transients in the instruments, whenever possible.
It is, however, non-trivial to do so because the origins of many of them are unknown or, in some cases, not in our control (e.g., earthquakes)~\cite{detchar-s6}. Those difficulties notwithstanding, detector characterization exercises and related data-analysis techniques have been employed for mitigating several kinds of noise transients in GW searches~\cite{noise-transients, Berger-2018, cornish:2014kda,Sukanta-2016-1,cavaglia:2018xjq,colgan:2019lyo,boudart:2022xib,mohanty:2023mjn}. In some of these cases the excess noise-power in the GW strain-data is due to bilinear or nonlinear couplings among various detector subsystems and components~\cite{ajith, Bose-2016, Da-Silva-Costa-2018,Bernardthesis, PhysRevResearch.2.033066, PhysRevD.101.042003}.
These include recent applications of machine learning algorithms such as NonSENS~\cite{PhysRevD.101.042003} and DeepClean~\cite{PhysRevResearch.2.033066}, which have 
been developed to variously regress linear, nonlinear, and nonstationary noise that typically last for tens of seconds or longer. For that purpose, certain ``witness" sensor data are used to estimate instrumental or environmental noise contributions to the GW
strain channel and regress them. These activities have demonstrated varying degrees of
success in lowering the strain noise in software and improving the sensitivity to astrophysical searches. 

In this work, we develop a technique for detecting the presence of nonlinear couplings at second order, i.e., quadratic nonlinearities, in the GW strain data. A third-order statistic that is useful for such identification is the {\em bicoherence}~\cite{Bose-2016}, specifically associated with non-Gaussian transients, or ``glitches". The traditional bicoherence~\cite{KimYC} construction, however, has certain limitations that can 
often result in misidentifcation of phase-coupled noise disturbances. To rectify it, we introduce the phase-randomized (PR) bicoherence analysis to GW data, and demonstrate its utility  with simulated signal studies in Sec.~\ref{sec:hos_qpc}. In practice, random noise can sometimes mimic the bicoherence response of phase-coupled (PC) disturbances, thereby leading to false detections, or false alarms, even when phase randomization is used.
To determine whether there is evidence of phase-coupling in the data or not, a hypothesis test is carried out between PC and non-phase-coupled (NPC) disturbances by employing the PR bicoherence analysis. Along the way, we derive an expression for the bicoherence value that minimizes the false identification of non-phase-coupled disturbances as phase-coupled ones
in Sec.~\ref{sec:det_threshold}. This is useful for assessing how significant any phase-coupled disturbance is based on its bicoherence value. In Sec.~\ref{app:App_gw_data}, we apply the PR auto-bicoherence analysis to GW data from LIGO Hanford and Livingston detectors, and the utility of this analysis to identify nonlinear couplings is explained in terms of three examples in real data.

\section{\label{sec:hos_qpc} Higher order statistics $-$ Quadratic Phase Coupling}

Higher-order statistics, such as the bicoherence at the $3^{rd}$ order, can be helpful in identifying the presence of nonlinear frequency and phase couplings in the data~\cite{Bose-2016}. The frequencies of different disturbances can nonlinearly combine to produce excess noise at other frequencies. The simplest and often the most dominant term in nonlinear couplings is the quadratic term, which results in the Quadratic Phase Coupling (QPC)~\cite{405097}. Take, e.g., a pair of disturbances or ``signals'', $x(t)$ and $y(t)$, that quadratically couple to form a new signal. Here we will be typically interested in a third time-series, $z(t)$, that is significantly influenced by that quadratic coupling, namely, 

\begin{align}\label{eq:quadratic_equation}
    z(t) = p ~ [~ x(t) + y(t) ~]^2 ~ + ~ q ~ [~ x(t) + y(t) ~] ~ + ~ C(t)\,,
\end{align}

\noindent where $p$ and $q$ are real numbers, and $C(t)$ is the noise. When $p=0$, one has $x(t)$ and $y(t)$ affecting $z(t)$ at most linearly. In such a case, the influence of $x(t)$ or $y(t)$ on $z(t)$ can be detected by computing the {\em coherence} of $x(t)$ or $y(t)$ with $z(t)$. However, when $p\neq 0$, the influence of $x(t)$ and $y(t)$ on $z(t)$ is nonlinear (quadratic, to be precise), and a third-order statistic, such as the bispectrum or bicoherence can be used to reveal this influence.

To compute the bicoherence statistic of these three time-series, we divide each of them
into $M$ concurrent segments, $x_ {\rm \alpha}(t)$, $y_{ \rm \alpha}(t)$ and $z_{\rm \alpha}(t)$, respectively, where $\alpha = 1,..., M$ is the segment index. Their Fourier transforms are $\tilde X_ {\rm \alpha}(f)$, $\tilde Y_{\rm \alpha}(f)$ and $\tilde Z_{\rm \alpha}(f)$, respectively. In practice, we will take the latter three ``functions'' as discrete Fourier transforms and their arguments as discrete frequencies. In terms of those transforms, the \textit{bispectrum} statistic is defined as

\begin{align}\label{eq:bispectrum}
  B (f_k, f_l) = \frac{1}{M} ~ \sum_{ \rm \alpha=1}^M ~ \tilde X_{\rm \alpha}(f_k) ~ \tilde Y_{\rm \alpha}(f_l) ~ \tilde Z_{\rm \alpha}^*(f_m)\,,
\end{align}
where
$f_k$, $f_l$, $f_m$ are taken to be discrete frequencies, with $k$ and $l=1,...,N$, 
$k\neq l $, and $ f_m = f_k \pm f_l $, such that $m \leq N $. Here, $N$ is the Nyquist frequency index and $f_N$ will denote the Nyquist frequency. Thus, the above frequencies obey the following limits: (1) $0 \leq f_{k,l}$ $\leq f_N$; (2) $ 0 \leq f_k \pm f_l \leq f_N$.
The bispectrum is computed for pairs of frequencies $(f_k,f_l)$ -- also called {\em bifrequencies} -- and obeys the following properties when $z_\alpha(t)$ is chosen to be the same as $y_\alpha(t)$, for all $\alpha$~\cite{Kim}: (1)  $B(f_k, f_l) = B^*(-f_k, -f_l)  = B(f_l, f_k)$; and (2) $B(f_k,f_l) = B(f_k, -f_k - f_l)$. If $\tilde{Z}$ has a signal that is phase-coupled with $\tilde{X}(f_k)$ and $\tilde{Y}(f_l)$, then the bispectrum can assume large values at the bifrequency $(f_k, f_l)$ if that coupling is strong. A normalized version of the  bispectrum is defined as:

\begin{align}\label{eq:bicoherence}
  b(f_k, f_l) = \frac{ \Big|B(f_k, f_l)\Big| } { \sqrt{E\Big[ \big| \tilde X(f_k) ~ \tilde Y(f_l) \big|^2 \Big]~  E\Big[ \big|\tilde Z(f_m)\big|^2 \Big]} }\,,
\end{align}
and is called the \textit{bicoherence}~\cite{Hagihira}. Here, $|\tilde{A}(f_l)|^2 \equiv 
\tilde{A}_{\rm \alpha}(f_l)\tilde{A}^*_{\rm \alpha}(f_l)$ and $E[ \tilde{A}(f_l) ] \equiv 
\frac{1}{M} ~ \sum_{ \rm \alpha=1}^M ~ \tilde{A}_{\rm \alpha}(f_l)$. Below, we will most often employ the bicoherence instead of the bispectrum. It is evident that the bicoherence obeys $0\leq b(f_k, f_l) \leq 1$. This is most readily seen by identifying the product
$\tilde X_{\rm \alpha}\tilde Y_{\rm \alpha}$ as the components of a single complex vector, say, $\tilde W$. It then follows that
\begin{align}
\frac{ B(f_k, f_l) } { \sqrt{E\Big[ \big| \tilde X(f_k) ~ \tilde Y(f_l) \big|^2 \Big]~  E\Big[ \big|\tilde Z(f_m)\big|^2 \Big]} }
=
\frac{ 
\sum_{ \rm \alpha=1}^M ~ \tilde W_{\rm \alpha} ~ \tilde Z_{\rm \alpha}^*}
{
\sqrt{\sum_{ \rm \alpha=1}^M ~ \left| \tilde W_{\rm \alpha}
\right|^2
}
\sqrt{\sum_{ \rm \alpha=1}^M ~ \left| \tilde Z_{\rm \alpha}
\right|^2}
}\,,\nonumber
\end{align}
which is just the inner product of two unit-norm complex vectors. The magnitude of such a product obeys the Cauchy-Schwarz inequality, which is just the constraint noted above on $b(f_k, f_l)$.

In the context of GW data analysis, $x(t)$, $y(t)$ and $z(t)$ are time-series data from the GW strain channel or any of the auxiliary GW detector channels (e.g., associated with instruments and sensors). When one sets $x(t) = y(t) = z(t)$ in Eq.~(\ref{eq:bicoherence}), the resulting statistic is called the \textit{auto$-$bicoherence}. Otherwise the statistic is also called the \textit{cross$-$bicoherence}.

The bispectrum is manifestly complex and its phase is called the \textit{biphase}, $\Phi_{\rm d}$. The variance of the biphase is denoted as $\sigma^{\prime~2}_{\rm ~\Phi}$. For the auto-bicoherence, $\Phi_{\rm d}$ is empirically related to it as~\cite{Elgar} 
\begin{align}\label{eq:sig_phi_sq}
\sigma^{\prime~2}_{\rm ~\Phi} \approx \frac{1}{2M} ~ \left[\frac{1}{b^2(f_k,f_l)}-1\right]\,,
\end{align}
and will be discussed further in the following section.On the other hand, if $\Phi^\alpha(f_k)$, $\Phi^\alpha(f_l)$, and $\Phi^\alpha(f_m)$ denote the phases of $\tilde X_ {\rm \alpha}(f_k)$, $\tilde Y_{\rm \alpha}(f_l)$ and $\tilde Z_{\rm \alpha}(f_m)$, respectively, then $\Phi^\alpha_{\rm d} \equiv \Phi^\alpha(f_k) + \Phi^\alpha(f_l) - \Phi^\alpha(f_m)$
is the biphase of individual segments. Given a bifrequency, if the corresponding biphases of 
the $M$ segments happen to be independent of one another, and have a uniform distribution in the interval $( -\pi, \pi]$, then its bispectrum  $\approx 0$. Its bicoherence will be negligible as well. On the other hand, when $\Phi_d^\alpha \approx 0$ for every segment,
then it will result in a non-vanishing bispectrum and a bicoherence $\approx 1$, which
is just the maximum value allowed by the Cauchy-Schwarz inequality.

\begin{figure}[htp!]
    \centering
		\includegraphics[scale=.5]{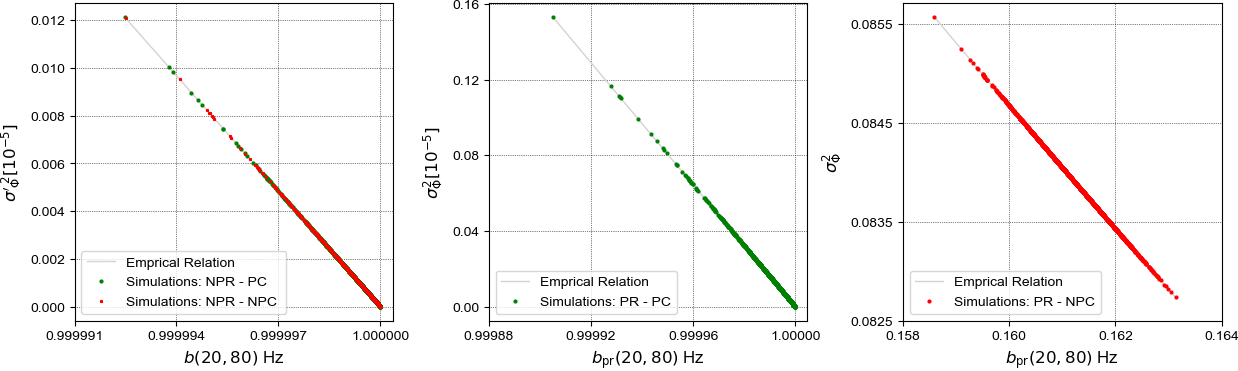}
\caption{The impact of phase randomization on distinguishing nonlinearly phase-coupled signals (labeled ``PC'' and shown as green dots) from those without (labeled ``NPC'' and shown as red dots), in plots of biphase variance {\em vs} (auto-)bicoherence. We compute these statistics for the $(20, 80)$ Hz bifrequency of the simulated PC and NPC signals described by Eq.~(\ref{eq:modeled_signal}). Each dot represents a different realization of the noise to which those signals are added.
{\it Left}: Here the {(auto-)bicoherence} is computed traditionally, i.e., with no phase randomization (NPR). Many PC signals show up with strong bicoherence values in this plot, but so do many NPC signals. The assumption that the bispectrum phases of the $M$ segments are independent of one another is invalid for these NPC signals and leads to their high bicoherence values, thus, making it difficult to distinguish them from the PC signals. 
{\it Middle}: The bicoherence values of the of PC signals shown in the left figure are minimally affected by the introduction of artificial phase randomization (PR).
{\it Right}: For the same NPC signals shown in the left figure, the use of phase randomization helps in identifying the lack of phase coupling in those cases and gets their bicoherence to stay low. (The corresponding biphase variance tends to rise.) This helps in distinguishing the NPC signals from the PC ones when phase randomization is applied.}
\label{fig:biphase_vs_bicoherence}
\end{figure}

Unfortunately, the bicoherence -- as {\em traditionally} defined in Eq.~(\ref{eq:bicoherence}) -- suffers from the limitation that it can turn out to be unity even when there is no phase coupling. This can happen, e.g., when the $\Phi_{\rm d}^\alpha$ are the same across all $M$ segments and, say, equal $\Phi_{\rm d}$. In that event, the bicoherence is just $|E[e^{i{\Phi_{\rm d}}}]| = |e^{i{\Phi_{\rm d}}}|$, which is unity. This would be a false alarm since that value is indistinguishable from the one expected of a phase-coupled signal, where the biphase is 0 and, consequently, the bicoherence is unity. One way to address this problem is to multiply the phase $\Phi_{\rm d}^\alpha$ of the data from each segment with a random variable $R^{(\rm \alpha)}$ such that $R^{ \rm (\alpha)}$ is the $\alpha$-th realization of a random variable that follows either a uniform distribution over $(-\pi,\pi]$ or a zero-mean normal distribution with a variance $\sigma^2_{\rm R}$. The phase-randomized bispectrum is:
%
\begin{align} \label{eq:phase_randomized_bispec}
B_{\rm pr}(f_k, f_l) = \frac{1}{M} ~ \sum_{\alpha=1}^M ~ \left| \tilde X_\alpha (f_k) ~ \tilde Y_\alpha (f_l) ~ \tilde Z_\alpha^{*}(f_m) \right |e^{iR^{(\rm \alpha)}\Phi_{\rm d}^\alpha}\,,
\end{align}
and the {\em phase-randomized} bicoherence $b_{ \rm pr}(f_k, f_l)$ is given by Eq.~(\ref{eq:bicoherence}) with $B(f_k, f_l)$ replaced by $B_{\rm pr}(f_k, f_l)$.

The PR prescription works equally well regardless of which of the two distributions is used for $R^\alpha$~\cite{Kim}. The expression for the variance of biphase corresponding to phase-randomized bicoherence is similar to Eq.~(\ref{eq:sig_phi_sq}). It is estimated by using $b_{ \rm pr}(f_k, f_l)$ instead of $b(f_k, f_l)$, and the variance of its biphase will be denoted as $\sigma_{\Phi}^2$.

The strength of the phase couplings in a signal at any particular bifrequency, $b_{ \rm pr}(f_k, f_l)$, is inferred from the bicoherence value, which lies between 0 and 1. A true nonlinearly coupled bifrequency has $b_{ \rm pr}(f_k, f_l) \approx 1$. For non$-$phase coupled bifrequencies, one has $b_{ \rm pr}(f_k, f_l) $ closer to zero, for large enough $M$ and when the noise in those segments is statistically independent. 

In this work, we focus on the auto$-$bicoherence and briefly present a feasiblity study on the cross$-$bicoherence in Appendix~C. The main reason for this decision is that the cross$-$bicoherence analysis is computationally expensive when {\it a priori} it is unclear which detector systems'  perturbations would be phase-coupling to produce specific noise transients in the GW strain channel. Current detectors log these data in hundreds of thousands of auxiliary and physical-environmental-monitor channels. Even if one were to pick the top hundred suspects from that list for $x$ and $y$, computing the cross$-$bicoherence with $x\neq y$, and $z$ as the GW strain, involves ${\cal O}(10^4)$ times more floating point operations than the auto$-$bicoherence, with all three data streams being the GW strain. Indeed, one can use certain well-motivated strategies, e.g., requiring the concurrence of transients in auxiliary channels with the GW strain data~\cite{smith:2011an} to draw up such a shortlist. A second idea is to use a two-step hierarchical strategy, where in the first step one computes the auto-bicoherence on the GW strain channel at the locations of the noise transients. In the second step, one evaluates the more computationally expensive cross-bicoherence but only for that subset of transients from the first step that return the largest auto-bicoherence values. Combining both ideas will help reduce both the number of channels and the noise transients and can make the computation of the cross-bicoherence manageable. Details of that study will be presented in a future work~\cite{bricewilliams}.
{For the rest of the paper, by {\em bicoherence} we will always mean the {\em auto$-$bicoherence}, unless otherwise noted.} 

Below we show with simulations how the bicoherence statistic can be used to discern the presence of quadratic phase coupling. Thereafter, we present some applications in real GW data. Throughout this paper, we fixed the values of some of the analysis  parameters for both simulated signals and GW data,  unless stated otherwise. The duration of all data chunks analyzed is chosen to be 4~sec, with a sampling frequency $f_{\rm s} = 1024$ Hz; of course, the Nyquist frequency is set to $f_N = \frac{f_{\rm s}}{2} $. For computing the bicoherence, each time-series data chunk is divided into $M$ segments and their Fourier transforms taken after applying the Hanning window on them.

The optimal choice for the number of segments ($M$) is decided empirically. For a fixed chunk duration, if $M$ is small, then the relative contribution of  noise to the bicoherence magnitude can be large. On the other hand, if $M$ is large, then the overlap of the segments will be large for the same chunk duration. In such a case, the different segments will cease to be independent and the bicoherence for a real signal, of a similar duration as the chunk, will plateau. By performing injection of simulated signals in real data and studying known short duration couplings (lasting no more than a few seconds) we found $M \approx 31$, which corresponds to a $98 \%$ overlap among every pair of neighboring segments.

With the above specifications, for each 4s chunk the auto-bicoherence computation takes ${\cal O}(M{\cal N}^3)$ floating-point operations (flop) with 8-byte long complex numbers,
where ${\cal N}= 4096$ frequency points. This adds up to about a tera-flop, which, along with read/write overheads, takes several minutes to execute on current processors. Thus, if the number of channels and noise transients can be kept manageable, the cross-bicoherence computation will be viable.

\subsection{Usefulness of the bicoherence: An example} 
We now setup a simulation of phase-coupled signals to demonstrate how the bicoherence can be used to detect that coupling. Consider the sinusoidal signals $x(t) = a_x \cos (2\pi f_x t+\Phi_x)$ and $y(t) = a_y \cos (2\pi f_y t+\Phi_y)$, with $a_{x,y}$, $f_{x,y}$ and $\Phi_{x,y}$ as the respective amplitudes, frequencies, and phases.  Using these signals
in Eq.~(\ref{eq:quadratic_equation}), and setting $q = 0$ and 
$C(t) = n(t) - p(a_x^2 + a_y^2)/2 $ there, we get:
%
\begin{align}\label{eq:quadratic_equation_simpleform}
    z(t) &= p ~ [~ a_x \cos (2\pi f_x t+\Phi_x) ~ + ~ a_y  \cos (2\pi f_y t+\Phi_y) ~]^2 ~ + n(t) - p\frac{(a^2_x + a^2_y)}{2}\,,
\end{align} 

\noindent where $n(t)$ is Gaussian noise with a vanishing mean. 
Expanding Eq.~(\ref{eq:quadratic_equation_simpleform}) by using trigonometric identities, the signal takes the form

\begin{figure}[htp!]
\centering
\includegraphics[scale=0.7]{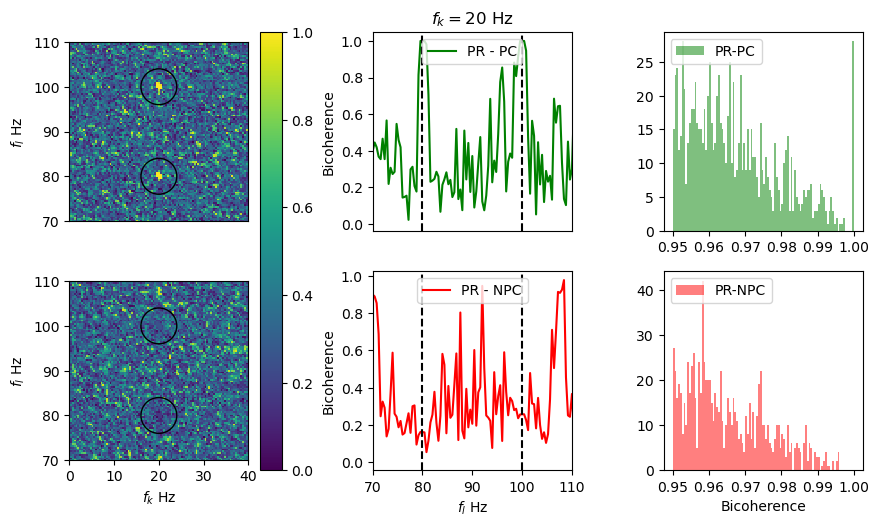}
\caption{The two circles in the top-left bicoherence map mark the two phase-coupled (PC) bifrequencies, namely, (20,80)~Hz and (20,100)~Hz, corresponding to the signal in Eq.~(\ref{eq:modeled_signal}), with bicoherence of $\simeq1$.  Contrastingly, in the bottom-left bicoherence map of the corresponding NPC signal, the same region shows a bicoherence of $\simeq0$, implying lack of phase coupling. Phase randomization (PR), which was used here, improves the identification of PC bifrequencies. To analyze the behavior of bicoherence at and around the bifrequencies $(20, 80)$~Hz and $(20,100)$~Hz, line plots are shown in the middle two panels, with $f_l \in [70 - 110]$~Hz and $f_k$ kept fixed at 20~Hz. The vertical dashed black lines correspond to the bifrequencies (20,80)~Hz and (20,100)~Hz. Top-middle: At the phase-coupled bifrequencies, the bicoherence is close to one. The bicoherence at those bifrequencies dips to low values for the non-phase-coupled case (bottom-middle plot). Note that for this (NPC) case, we chose $\Phi_{\rm zp} = 0 = \Phi_{\rm zn}$. The right panel shows the bicoherence histogram of the left panel. The bottom-right figure shows that the probability of misclassifying an NPC as PC signal is 7e-4 for $b_{\rm pr} > 0.995$, but drops to 0\% above a value of $b_{\rm pr} = 0.996$ (in this experiment).
}
\label{fig:bicoherence_simsignal}
\end{figure}

\begin{align}\label{eq:modeled_signal}
z(t)  &= ~ p \left[~ \frac{a_x^2}{2} ~ \cos(4\pi f_x t + 2\Phi_x) ~ \right] ~ + ~ p \left[~ \frac{a_y^2}{2} ~ \cos(4 \pi f_y t + 2\Phi_y) ~ \right] \noQ 
& + ~ p \left[~ a_x a_y ~ \left\{~ \cos( 2\pi f_{\rm zp} t+ \Phi_{\rm zp} ) ~+~ \cos( 2\pi f_{\rm zn} t + \Phi_{\rm zn} ) ~ \right\} ~ \right]  ~ + ~ n(t)\,,
\end{align} 

\noindent 
where 
$f_{\rm zp} \equiv (f_x + f_y)$, 
$f_{\rm zn} \equiv (f_x - f_y)$,
$\Phi_{\rm zp} \equiv \Phi_x + \Phi_y$ 
and $\Phi_{\rm zn} \equiv \Phi_x - \Phi_y$. 

In our simulations we constructed two kinds of signals:
(a) A phase-coupled (PC) signal, modelled after $z(t)$, where $\Phi_{\rm zp}$ and
$\Phi_{\rm zn}$ obey the preceding relations with 
$\Phi_x$ and $\Phi_y$; 
(b) A non-phase-coupled (NPC) signal, 
which is simulated similar to $z(t)$, except that it has 
$\Phi_{\rm zp} \neq \Phi_x + \Phi_y$ and $\Phi_{\rm zn} \neq \Phi_x - \Phi_y$.
In fact, we chose the input parameter values in the simulated signals such that $f_x = 40$ Hz, 
$f_y = 60$ Hz, $\Phi_x = 40^{\circ} $, $\Phi_y = 60^{\circ}$, and
$a_x = a_y = \sqrt{2} $. Also, the standard deviation of $n(t)$ was taken to be
much smaller than $p(=1)$ and was, in fact, set to 0.0025.
For the PC signal, 
$f_{\rm zp} =100$ Hz,  
$f_{\rm zn} = -20$ Hz,
$\Phi_{\rm zp} = 100^\circ$ and
$\Phi_{\rm zn} = -20^\circ$. Whereas for the NPC signal, all parameters were set to the same values as the PC signal except that 
$\Phi_{\rm zp} = 0=\Phi_{\rm zn}$.

If one were to compute the cross-bicoherence of $x(t)$, $y(t)$, and $z(t)$, then one can expect high bicoherence values at  the bifrequency of $(40,60)$~Hz, since signals at these frequencies are present in $x$ and $y$, respectively, and the sum and difference (actually the absolute value of the difference)  of those frequencies, namely, 100~Hz and 20~Hz, are represented in $z$ as $f_{\rm zp}$ and $f_{\rm zn}$, respectively.

However, when computing the auto-bicoherence of $z$, note that the four frequencies
present in it are $2f_x = 80$ Hz, $2f_y = 120$ Hz, $|f_{\rm zn}| = 20$ Hz and $f_{\rm zp} = 100$ Hz. The only independent bifrequencies chosen from this set whose sum(s) or difference(s) can be found in the same set are $(20,80)$~Hz (since 100 Hz is present in the same set) and $(20,100)$~Hz (since 120 Hz is also present in the same set). This is, of course, expected, given the origin of $z$ in $x$ and $y$. (Note that the ``reflected'' pairs $(80,20)$~Hz and $(100,20)$~Hz will also have strong bicoherence values but are not independent of the above pairs, as described in Appendix~A.)

Given the importance of the bicoherence maps in this work, it is worthwhile to go over this logic in more detail: In the example of Eq.~(\ref{eq:modeled_signal}), $z(t)$, or its Fourier transform $\tilde{Z}$, has power in four frequencies, namely, $2f_x = 80$~Hz, $2f_y = 120$~Hz,
$f_{\rm zp} =(60+40){\; \rm Hz} = 100$~Hz and $-f_{\rm zn} =(60 - 40){\; \rm Hz}=20$~Hz (keeping to positive frequency values). This will remain true for all $M$ segments constructed from the original time-series $z(t)$ by following the procedure dsrcribed above.
As can be established by setting $\tilde{X}_\alpha = \tilde{Y}_\alpha \equiv \tilde{Z}_\alpha$, for all $\alpha$, in the definition of the bispectrum in Eq.~(\ref{eq:phase_randomized_bispec}), one of the three conditions that is necessary but not sufficient for it, and the associated auto-bicoherence, to assume large values for any bifrequency is that $f_m = f_l + f_k$. For the current example, $f_l$ and $f_k$ can be
any pair chosen from the same set of four frequencies, {\it viz.}, $\{20, 80, 100, 120\}$~Hz, 
but the bispectrum will be large only when $f_m$ also assumes one of the same four values. This is the second necessary condition. Consequently, when $f_l=20{\;\rm Hz}=f_k$, one has $f_m=40$~Hz, which, however, is not present in the above set of four frequencies and, hence, one does not find strong bispectrum, or bicoherence, at the bifrequency of (20,20)~Hz. 
For similar reasons, the bifrequency of (80,80)~Hz also does not have a strong bicoherence value.

Contrastingly, with $f_l=20$~Hz and $f_k=80$~Hz, one has $f_m=100$~Hz, which is
present in the set of four. Hence, the bifrequency (20,80)~Hz satisfies the first two conditions. The third and final condition that must be satisfied is that $\Phi_{\rm d}^\alpha \equiv \Phi^\alpha(f_k) + \Phi^\alpha(f_l) - \Phi^\alpha(f_m)$, defined above Eq.~(\ref{eq:sig_phi_sq}), must vanish at those frequencies, for each one of the $M$ segments.
In this example, the phase-coupled signal obeys, for all segments, $\Phi_{\rm d}^\alpha = 
\Phi^\alpha({\rm 20~Hz}) + \Phi^\alpha({\rm 80~Hz}) - \Phi^\alpha({\rm 100~Hz})
= -\Phi_{\rm zn} + 2 \Phi_x - \Phi_{\rm zp} = -(\Phi_x - \Phi_y) + 2\Phi_x - (\Phi_x + \Phi_y)$, which vanishes regardless of the specific values that $\Phi_x$ and $\Phi_y$ may take in a given realization of the (quadratic) phase-coupling in Eq.~(\ref{eq:modeled_signal}). Thus, the signal $z(t)$ in this example obeys all three conditions for the bifrequency (20,80)~Hz and, consequently, its bicoherence assumes a near unity value at that bifrequency. Note that the noise $n(t)$ will also contribute a random element to $\Phi_{\rm d}^\alpha$. However, when taking the average over a sufficiently large number of segments $M$ its influence will be small. For true phase-coupled signals present, like in Eq.~(\ref{eq:modeled_signal}), their contribution to the biphase $\Phi_{\rm d}^\alpha$ will be consistently zero for all $M$ segments, which helps the bicoherence attain a value close to unity.

Similarly, with $f_l=20$~Hz and $f_k=100$~Hz, one has $f_m=120$~Hz, which is present in the set of four. Hence, it too receives strong bispectrum and bicoherence values. Owing to the symmetry properties of the bispectrum discussed earlier the bifrequencies (80,20)~Hz and (100,20)~Hz will also receive large values for those statistics. Note, however, that  bispectrum / bicoherence maps often do not display these regions of redundant bifrequencies (see \ref{app:App-D} for more details). One can likewise show that the remaining bifrequency pairs constructed from the aforementioned set of four frequencies will not receive significant bicoherence values.

To test the above expectations about the bicoherence results for the above example through simulations, we added the PC and NPC versions of $z(t)$, separately, into 2000 realizations of noise, $n(t)$. For these simulated data sets, for PC and NPC cases, we then computed the (auto-)bicoherence values and their corresponding biphase variances without applying phase randomization. We repeated the experiment after applying phase randomization, and plotted results from both studies in Fig.~\ref{fig:biphase_vs_bicoherence}, which shows that phase randomization improves the distinguishability of phase-coupled signals from non-phase-coupled signals. 

The bicoherence maps of the same simulated signals, for the PC and NPC cases, are shown in Fig.~\ref{fig:bicoherence_simsignal} (left panel). The right panel in Fig.~\ref{fig:bicoherence_simsignal} is the non-redundant~\footnote[1]{See \ref{app:App-D}.} bifrequencies histogram. The middle panel shows the values of $b_{\rm pr}$ as a function of $f_l$ when $f_k$ is held fixed at 20~Hz. The bottom-right figure shows that the probability of misclassifying an NPC as PC signal is 7e-4 when the  bicoherence ($b_{\rm pr}$) threshold is set to 0.995, but drops to 0\%  for $b_{\rm pr} > 0.996$ (in this experiment). The non-linear phase couplings in the signal are identified correctly through the phase randomized bicoherence method. In this paper, the bicoherence maps are positive quadrants of the bicoherence plane (see \ref{app:App-D}). Note that each octant in the bicoherence plane captures complete information about the bifrequencies owing to its eight-fold symmetry.

\section{\label{sec:det_threshold} Estimating detection threshold}

In the previous section, we demonstrated that applying phase randomization
in the computation of bicoherence improves the distinguishability of PC and NPC cases.
This improvement rests on the following reasonable assumptions about the signals: (i) the phase component of bifrequencies, $\Phi_k^\alpha$ and $\Phi_l^\alpha$, are uniformly distributed random variables, over  $(-\pi,\pi]$; (ii.a) For the NPC bifrequencies, we shall take that $ \Phi_m^\alpha \neq 0 \neq \Phi_{\rm d}^\alpha$; (ii.b) For NPC cases, $ \Phi_m^\alpha $ and $ \Phi_{\rm d}^\alpha$ are uniformly distributed as $\sim U(-\pi,\pi]$~\cite{Kim} and independent of $\Phi_k^\alpha$ and $\Phi_l^\alpha$. The worst case scenario, as discussed above, is when all $\Phi_d^\alpha$ are identical, say, $\Phi_d$, which falsely produces large bicoherence values and necessitated the use of phase randomization;
(iii) For the PC bifrequencies, $ \Phi_{\rm d}^\alpha = 0 $ for all $\alpha$ -- although, in practice, they are not exactly zero and can be assumed to have a normal distribution, i.e., $\Phi_{\rm d} \sim N(0,\sigma_{\Phi}^2)$~\cite{Elgar}. However, the true phase-coupled bifrequencies can become indistinguishable from noise when the signal's power or the strength of the phase coupling is low. This can lead to a false dismissal of PC. Therefore, a test is required to decide whether or not there is phase coupling present at any bifrequency. 
We assume two hypotheses that are tested on the simulated data:

\begin{align*}
{\rm Null ~ hypothesis, ~ H}_0 &: ~ {\rm NPC}, ~ \Phi_{\rm d} \sim U(-\pi,\pi], ~ ~ b_{\rm pr}^2 (f_k, f_l) \approx 0  \,. \noQ 
{\rm Alternative ~ hypothesis, ~ H}_1 &: ~ {\rm PC}, ~ ~ \Phi_{\rm d} \sim N(0,\sigma_{\Phi}^2), ~ ~ b_{\rm pr}^2(f_k, f_l) \approx 1 \,. \noQ
\end{align*}
We define $T$ as the detection  threshold for bicoherence. The two types of errors associated with these  hypotheses and the choice of $T$ that we are interested in are the false-positive and false-negative errors~\cite{Kim}. Suppose in reality there is no  phase-coupled signal in the data. In such a case, finding $b_{\rm pr}^2(f_k, f_l) > T$ implies a false detection of PC. This is a false-positive error and its probability of occurrence is: 

\begin{align}\label{eq:prob_fp1}
P_{\rm {fp}} &= P ~ \{ \rm {H}_1 ~|~ \rm {H}_0 \} \noQ
       &= P ~ \left \{ b_{\rm pr}^2 (f_k, f_l) ~ > T ~|~ \rm{NPC} \right \}\,.
\end{align}

\noindent Using Eq.~(\ref{eq:phase_randomized_bispec}) one can show that the phase randomized bicoherence obeys $b_{\rm pr}^2(f_k, f_l) = e^{- \sigma_{\rm R}^2 {\Phi_{\rm d}}^2}$, where 
the randomization variable $R^\alpha$ for every segment was chosen to be distributed as $ N (0, \sigma_{\rm R}^2)$ for mathematical convenience~\cite{Kim}. Then, Eq.~(\ref{eq:prob_fp1}) becomes

\begin{align}\label{eq:prob_fp}
P_{\rm {fp}} &= P ~ \left \{ e^{- \sigma_{\rm R}^2 \Phi_{\rm d}^2} ~ > T ~| ~ \Phi_{\rm d} \sim U(-\pi,\pi] \right \} \noQ
       &= P ~ \left \{ |\Phi_{\rm d} | < ~ \frac{\sqrt{ { \rm ln} \frac{1}{T} }}{\sigma_{\rm R}} ~|~ \Phi_{\rm d} \sim U(-\pi,\pi] \right \} \noQ
       & = \frac{\sqrt{ {\rm ln} \frac{1}{T} } }{\pi \sigma_{\rm R}} \,,
\end{align}

\noindent which shows that as $T$ increase the probability of false$-$positives decreases. On the other hand, when the phase-coupled bifrequency has bicoherence value less than the threshold value, i.e., $b_{\rm pr}^2(f_k, f_l) \leq T$, one has a case of false rejection of PC. The probability of occurrence of such false$-$negatives is

\begin{align}\label{eq:prob_fn}
P_{\rm fn} & = P ~ \{ {\rm H}_0 ~|~ {\rm H}_1 \} \noQ
       & = P ~ \left \{ b_{\rm pr}^2 (f_k, f_l) ~ \leq ~ T ~|~ \rm {PC} \right \} \,.
\end{align}

\noindent As discussed before, in  practical cases of PC signals, one has $\Phi_{\rm d} \approx 0$, with a small variance. In fact, it is normally distributed as $\approx N(0, \sigma_{\rm \Phi}^2)$, with $\sigma_{\rm \Phi}$ characterizing the phase coupling~\cite{Elgar}. With this input, Eq.~(\ref{eq:prob_fn}) becomes

\begin{align}\label{eq:prob_fn_exp}
P_{\rm fn} & = P ~ \left \{ e^{- \sigma_{\rm R}^2 {\Phi_{\rm d}}^2} ~ \leq T ~| ~ \Phi_{\rm d} \sim \emph{N}(0,\sigma^2_{\rm \Phi}) \right \} \noQ
       & = P ~ \left \{ |\Phi_{\rm d} | ~ \geq ~ \frac{\sqrt{ {\rm ln} \frac{1}{T} }}{\sigma_{\rm R}}  ~| ~ \Phi_{\rm d} \sim \emph{N}(0,\sigma^2_{\rm \Phi}) \right \} \noQ
       & = 2\int_{\lambda}  ^{\infty}  N e^{- \frac{{\Phi_{\rm d}}^2}{2\sigma^2_{\rm \Phi}}} ~ d\Phi_{\rm d} \,,
\end{align}
where $\lambda =  \frac{\sqrt{ {\rm ln} \frac{1}{T} }}{\sigma_{\rm R}} $ and $ N = \frac{1}{\sqrt{2\pi} \sigma_{\Phi}}$. 

The bicoherence value, $T_{\rm min}$,  that minimizes  the probability of detection error, $P_{\rm E} = P_{\rm fp} + P_{\rm fn}$, can be estimated by differentiating $P_{\rm E}$ with respect to $T$~\cite{Kim}; it is found to be:

\begin{align}\label{eq:min_threshold}
T_{\rm min} = \bigg(\frac{\sigma_{\rm \Phi}}{\sqrt{2 \pi}}\bigg)^{2 \sigma_{\rm \Phi}^2 \sigma_{\rm R}^2}\,.
\end{align}
Typically, $T_{\rm min}$ is set as the lower limit of the phase-randomized bicoherence for identifying cases of phase-coupled signals. Substituting $T_{\rm min}$ in Eqs.~(\ref{eq:prob_fp}) and (\ref{eq:prob_fn}), we get $P_{\rm fp}$ and $P_{\rm fn}$, which now depend on $\sigma_{\Phi}$ alone. Indeed, $P_{\rm fp}$ can be expressed as 
%
\begin{align}\label{eq:min_prob_fp}
P_{\rm fp} &= \frac{\sigma_{\rm \Phi}}{\pi} \sqrt{ {\rm ln} \left (~\frac{2\pi}{\sigma_{\rm \Phi}^2} \right)}\,.
\end{align}
The expression for the corresponding $P_{\rm fn}$ remains the one in Eq.~(\ref{eq:prob_fn_exp}), with the lower limit of that integration now set to $\lambda = \sigma_{\Phi } \sqrt{\rm ln ~\left(\frac{2\pi}{\sigma_{\rm \Phi }^2} \right)}$. The $P_{\rm fp}$ and $P_{\rm fn}$ corresponding to  $T_{\rm min}$ present the advantage that they can be estimated  directly from the variance of the biphase, and are cheaper to compute in practice.

\begin{figure}[htp!]
	\centering
    \includegraphics[scale=.72]{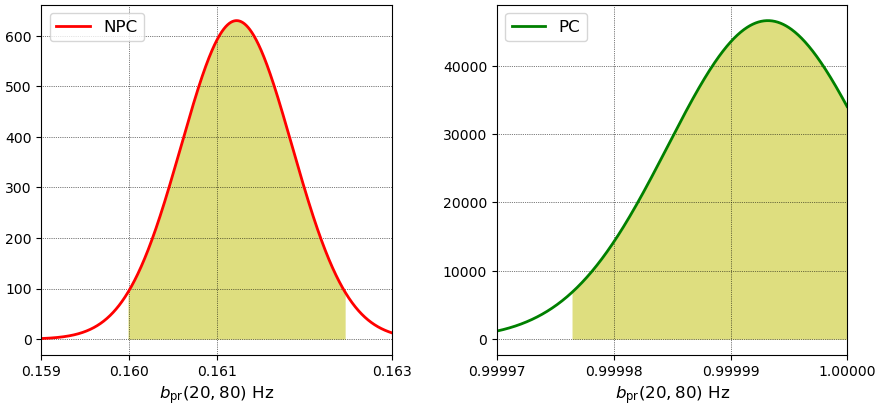}
	\caption{Distributions of the bicoherence values of the simulated non-phase coupled (NPC) signals (left) and the phase-coupled (PC) ones (right) described in Eq.~(\ref{eq:modeled_signal}), at the bifrequency (20,80)~Hz. Each shaded region corresponds to a $95\%$ confidence interval. These regions have very little overlap. 
 Therefore, one can afford to set the bicoherence threshold to be quite high in this simulation, e.g., at $b_{\rm pr} = 0.9999$, and benefit from a low false-negative probability while also enjoying a low false-positive probability.}
	\label{fig:distributions_3plots_vertical}
\end{figure}

The two hypotheses were tested on the simulated data sets. One can anticipate that the bicoherence value of PR$-$PC is closer to one, whereas PR$-$NPC is closer to zero. This anticipation is true in NPC and PC distributions of simulated signals at $b_{\rm pr}(20, 80)$~Hz, as shown in Fig.~\ref{fig:distributions_3plots_vertical}. The shaded regions in both plots of Fig.~\ref{fig:distributions_3plots_vertical}  each correspond to a 95\% confidence interval. These regions are well separated, and the probability of false-positive is low even for a high value of the bicoherence threshold. In real data applications below, we choose to be conservative in identifying phase-coupled signals, and set the threshold of the bicoherence {$b_{\rm pr}$} to be high, {\it viz.} {$b_{\rm pr}=0.98$}. The false-alarm probability at such high values is quite low, as evidenced in Fig.~\ref{fig:biphase_vs_bicoherence}.~\footnote[2]{As we explain below, the false-alarm probability is further lowered owing to the fact that we apply a threshold on the strength of the transient non-Gaussian noise chosen for follow-up search for non-linear couplings.}
Consequently, we analyze only the most significant cases of suspected PCs. As we make advances in developing more computationally efficient algorithms for computing bicoherence, we will lower this threshold further. 


\section{\label{app:App_gw_data} Application to gravitational-wave data }

In this section, we study  applications of  bicoherence for finding phase-coupled signals in real GW data. Certain disturbances in detector systems or their environment (e.g., due to scattered light or seismic vibrations) can sometimes introduce excess  noise in the detector strain and affect the quality of GW strain data. In this work, we are interested in disturbances that create transient non-Gaussian noise  -- also called {\em glitches}, which typically last for a few milliseconds to several seconds~\cite{Abbott-2016, Zevin-2017}. Certain classes of glitches (e.g., blips) are known to trigger  templates used for searching GW signals from compact binary coalescences (CBCs), thereby creating false alarms. The origin of several classes of glitches is not entirely known. Some others are known to originate through linear or non-linear coupling mechanisms among auxiliary channels~\cite{Bose-2016, Da-Silva-Costa-2018}. The auto-bicoherence can be an initial diagnostic tool for identifying 
potentially phase-coupled glitches in the GW strain data. This work focuses on such an  analysis in open-source data~\cite{GWOSC}.

\begin{figure}[htp!]
	\centering
	\includegraphics[scale=0.65]{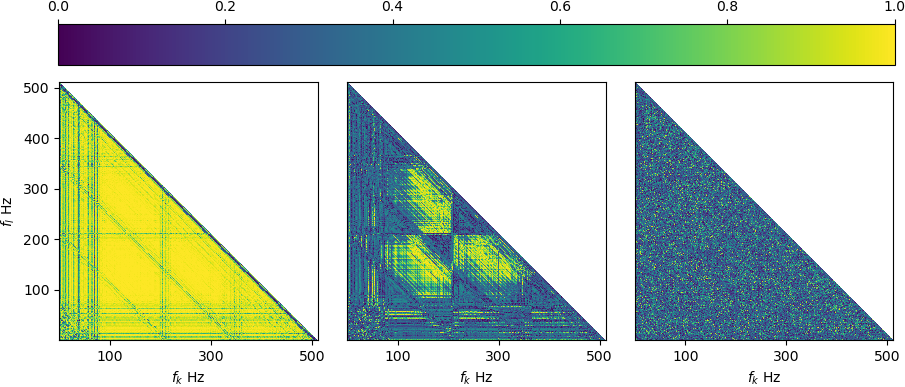}
	\caption{The bicoherence map of a blip glitch, constructed without phase randomization (PR), is shown on the left. The colorbar shows the bicoherence. It incorrectly shows many coupled bifrequencies, which disappear when the bicoherence is computed using phase randomization, as shown in the middle figure. This figure also reveals that the glitch may have been triggered from disturbances in frequencies that up-converted due to phase couplings. \textit{Column-3:} For comparison, we show here the PR bicoherence map of the background noise data that does not contain any glitch. Such data show the absence of any significant phase coupling. {Without the application of PR to this glitch-less data snippet, the bicoherence map (not shown) has numerous points with  bicoherence $\approx 1$  strewn across the map.}}
	\label{fig: prnpr_cmp}
\end{figure}

Importantly, failing to apply artificial phase randomization in the computation of auto-bicoherence can lead to false detection of PC bifrequencies. Consequently, the bicoherence maps also show different patterns depending on whether PR is used or not. One of the blip glitches that occurred in a LIGO Hanford detector during the O2 observation run with high excess power is demonstrated in Figure~\ref{fig: prnpr_cmp}, where the effect of PR can be clearly discerned. We continue to apply PR in all our bicoherence analysis below.

\subsection{\label{subsec:2dvs3d} Using bicoherence to classify glitches}

Instances of transient excess noise often show up in auxiliary  channels and are detected 
by tools such as Omicron~\cite{Robinet-2020}.~\footnote{{Also see Ref.~\cite{boudart:2022xib} for a machine learning based approach to noise transient detection.}}  
This tool performs a multi-resolution time-frequency analysis of detector data around such instances. Glitches with excess power above a given threshold are flagged for further inspection in data quality studies and astrophysical searches. Also, the time-frequency (TF) maps or spectrograms of Omicron serve to classify glitches via human intervention and machine learning ~\cite{Zevin-2017, Glanzer_2023}.

The blip glitches~\cite{Cabero-2019} form one of the categories of glitches. These are short noise transients that have a high-frequency bandwidth and often trigger CBC templates. 
In fact, it is the high-mass binary black hole search that is affected most by them. This is because these signals have a small number of wave-cycles in band and are of durations
comparable  to those glitches~\cite{joshi}. Blips appeared at the rate of approximately 4 per hour at Livingston (L1)  and 2 per hour at  Hanford (H1) during their third observation run (O3)~\cite{Davis-2021}. Koi fish and Tomtes are two other classes of glitches. While their morphologies share some characteristics with blips, important distinctions exist (see, e.g., Refs.~\cite{Cabero-2019, 41263, Ruxandra-2023}). Extremely loud glitches~\cite{Davis-2021} (with Omicron SNR $> 100$)  are another class of glitches that adversely affect a detector's sensitivity to searches of transient astrophysical signals. Studies have been carried out to identify the origin of these glitches, but no clear conclusion has been arrived at yet for several glitch types.

\begin{figure}[htp!]
	\centering
    \includegraphics[scale=.74]{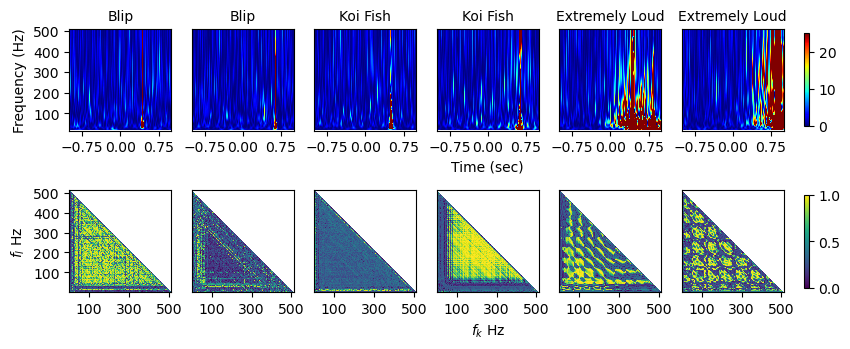}
    \caption{A comparison of second-order (top row) and third-order (bottom row) statistics of glitches. The top row shows the spectrograms of pairs of blips (columns 1-2), koi fish (columns 3-4) and ``extremely loud'' glitches (columns 5-6). The spectrograms look similar within each category, e.g., blips, simply because the glitches were categorized by the patterns revealed in those maps in the first place. They are different -- to varying degrees -- across categories, e.g., blips and extremely loud glitches. The bottom row shows the corresponding phase-randomized bicoherence maps, constructed using  4~sec data chunks centered at the glitches. Variations in the bicoherence maps within some glitch categories (e.g., columns 3 and 4 of koi fish) suggest the presence of different phase-coupled bifrequencies or a difference in the nature of that coupling. On the other hand, similarities in these maps across glitch categories (e.g., columns 1 and 4) suggest the presence of similar nonlinear phase couplings in different glitch types. Information in both types of maps, combined, can help in troubleshooting the origin of these noise artifacts.}
    \label{fig:3rdorder}
\end{figure}

The bicoherence maps of the blip, koi fish, and extremely loud glitches are produced by applying the phase-randomized auto-bicoherence analysis to look for $3^{\rm rd}$-order similarities in each class. If the observed glitches have a common source, their bicoherence maps will have similarities owing to common bifrequencies related to that source. Such expectations have been borne out in the study of $2^{\rm nd}$-order statistics, such as the TF maps of Omicron, as depicted in the upper panel of Fig.~\ref{fig:3rdorder}. The top row in that figure shows the spectrograms of pairs of blips (columns 1-2), koi fish (columns 3-4) and ``extremely loud'' glitches (columns 5-6). The spectrograms look similar within each category, e.g., blips, simply because the glitches were categorized by the patterns revealed in those maps. They are different -- to varying degrees -- across categories (e.g., blips and extremely loud glitches). The bottom row shows the corresponding phase-randomized bicoherence maps, constructed using  4sec data chunks centered at the glitches. Variations in the bicoherence maps within some glitch categories (e.g., columns 3 and 4 of koi fish) suggest the presence of different phase-coupled bifrequencies or a difference in the nature of that coupling. On the other hand, similarities in these maps across glitch categories (e.g., columns 1 and 4) suggest the presence of similar nonlinear phase couplings in different glitch types. (A few more examples of glitches with similar morphology in the third-order statistic are presented in~\ref{app:App-E}.) These maps indicate, e.g., that all blips are not made identically. In future, cross-bicoherence maps can help diagnose their origins by revealing what detector systems couple non-linearly to produce them.

It is also interesting to inquire what bicoherence maps look like for data with short-duration
Compact Binary Coalescence (CBC) signals. We leave this question for a future work to investigate but provide some preliminary findings in \ref{app:C1}.

\subsection{\label{subsec:recurrent} Recurrent bifrequencies in the quieter Omicron tiggers}

\begin{figure}[htp!]
    \centering
        \includegraphics[scale=.62]{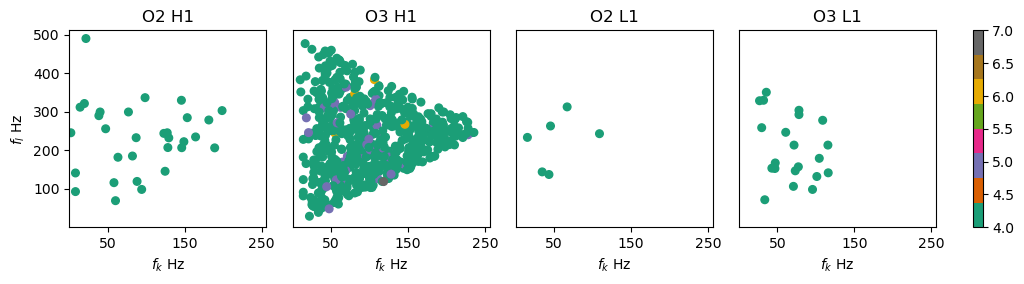}
        \caption{Recurrent bifrequencies with $ b_{\rm pr} \geq 0.98$  among the quieter Omicron triggers. When triggers with the same bifrequency occur more than three times during a run, we record them. These bifrequencies, from the second and third observation runs of H1 and L1 detectors, are shown here. The colorbar represents the number of such occurrences.}
        \label{fig:bifrequency_couplings}
\end{figure}

In this subsection, we report on results from the PR auto-bicoherence analysis performed to identify repeated phase-coupled bifrequencies in the relatively quiet Omicron triggers in O3. These triggers have received inadequate attention~\cite{Mogushi_2021} and can be detrimental to the quality of GW data. Every quiet omicron trigger chosen for bicoherence analysis here satisfies two conditions: (1) The Omicron SNR lies in the narrow range $[5,5.001]$;
(2) No other Omicron trigger exists within $\pm 3$~sec duration of the trigger time of interest. Based on these criteria, we selected the 
triggers shown in Fig.~\ref{fig:bifrequency_couplings}.

\begin{table}[ht!]
	\centering
	\footnotesize
	\begin{tabular}{|m{4.0cm}|m{4cm}|m{4cm} |}  \hline
			\textbf{Observation Run} & \textbf{Hanford (H1)}  & \textbf{Livingston (L1)} \\ \hline
                O2 & 751 & 534 \\ \hline
                O3 & 1079 & 440 \\ \hline
	\end{tabular}
	\caption{The number of ``quiet'' Omicron triggers considered here for identifying  recurrent bifrequencies in the O2 and O3 observation runs.}
	\label{tab:ntriggers}
\end{table}

In Fig.~\ref{fig:bifrequency_couplings}, we plot the bifrequencies of those quiet triggers that have $b_{\rm pr}\geq 0.98$. This value is somewhat less than the bicoherence threshold discussed for simulations in earlier sections but not too small to increase the false-positive rate significantly. When the same PC bifrequencies are found in more than three quiet triggers in the same detector, then that type of coupling is noted here as significant. These  select bifrequencies, from the O2 and O3 observation runs of L1 and H1 detectors, are displayed in Fig.~\ref{fig:bifrequency_couplings}. To probe possible causal connections with known noisy instances, the sum of the frequencies in each bifrequency pair in this shortlist is compared with the frequencies of gated lines~\cite{O2H1-lines, O3H1-lines, O2L1-lines, O3L1-lines} in those observation runs.~\footnote[3]{Differences of frequencies in some bicoherence triggers have also been found to be associated with gated lines but are not listed here, even though investigation continues on the origin of those features.} 
When that comparison reveals a match, an instance of non-linear coupling can be suspected. 
The closest matches are reported in Table~\ref{tab:noise_pairs_d98}. This comparison does not reveal any significant match in the L1 detector. The H1 detector, however, shows some interesting cases that merit further study to establish causal connections. While the auto-bicoherence does not establish, on its own, the presence of nonlinear frequency couplings or their cause, it does help in narrowing down the broad sensitivity band of the detectors to a manageably short list of frequencies that can be followed-up with additional analysis, such as an examination of auxiliary channels with the {\em cross-bicoherence}, for that purpose.

We also note that the nonlinear couplings between the mechanical resonances of test masses reported in our earlier work~\cite{Bose-2016} are found to exist in more recent observation runs as well. Specifically, the bounce and the violin modes of test masses (which register in the bicoherence map at bifrequency (10, 495.2) Hz)  as well as the power line harmonics and their sub-harmonics  identified in that analysis can be found in O3 too.

\begin{table}[ht!]
	\centering
	\footnotesize

	\begin{tabular}{|m{3.0cm}|m{3.5cm}|m{2.5cm}|m{5cm} |}  \hline
		\textbf{Bifrequency} & \textbf{Bifrequency sum} &
		\textbf{Gated lines} &
		\textbf{Gated lines} \\
		 $b_{\rm pr}$($f_k, f_l$) Hz & (Hz) & (Hz) & Comments \\ \hline

    \multicolumn{4}{|c|}{\cellcolor{lightcol} O2 H1} \\ \hline
    
     198.4, 303.2 &  501.6  & 501.61 & Violin mode 1st harmonic ITMY Mode 1 \\ \cline{3-4}
 		          &         & 501.68 & Violin mode 1st harmonic ITMY Mode 2 \\ \hline

	 \multicolumn{4}{|c|}{\cellcolor{lightcol} O3 H1} \\ \hline
     10.0,  383.2 & 393.2 & 393.2 & Calibration line non-linearity \\  \hline        
     16.4,  477.2 & 493.6 & 493.61514 & Likely calibration line mixing \\
	              &       & 493.62598 & 
               \\
	              &       & 493.62653 & \\	\hline	 
	 30.0,  54.4  & 84.4  & 84.4  & Calibration line non-linearity \\ \cline{1-3}
	 38.0,  66.4  & 104.4 & 104.4 & \\ \hline
	 
	 40.0,  448.8 & 488.8 & 488.78625 & Likely calibration line mixing \\ \hline
	 
	 48.8,  250.8 & 299.6 & 299.6 & Beam splitter violin mode region \\ \cline{1-1}
	 55.2,  244.4 &  &  &  \\ \hline
	 68.4,  420.4 & 488.8 & 488.78625 & Likely calibration line mixing\\ \hline	  
	 78.0,  423.6 & 501.6 & 501.63223 & Violin mode 1st harmonic\\
	              &       & 501.69513 & \\ \hline
	 134.0, 337.6 & 471.6 & 471.68625 & Likely calibration line mixing \\ \hline
	 150.8, 242.4 & 393.2 & 393.2     & Calibration line non-linearity \\ \hline
	 155.6, 333.2 & 488.6 & 488.78625 & Likely calibration line mixing \\ \cline{1-3}
	 206.0, 282.4 & 488.4 & 488.48624 &  \\ \hline
	 221.2, 287.6 & 508.8 & 508.84515 & Violin mode 1st harmonic\\ \hline
	\end{tabular}
	\caption{A subset of gated lines in H1, from O2 and O3, that are associated with high bicoherence. Only some of the bifrequencies  with $b_{\rm pr} \geq 0.98$ are listed in column 1, for illustration. The sum of the frequencies in a bifrequency is listed in column 2. Column 3 lists the subset of gated line frequencies in H1, from the same two observation runs, that turn out to be closest to the bifrequency sum listed in column 2. No such association was found in L1 (in O2 and O3) at the same level of significance.
It is worth noting that there are additional bifrequencies in those H1 and L1  datasets with equally or more significant bicoherence values whose bifrequency sums do not match any gated line frequencies in that data.}
	\label{tab:noise_pairs_d98}
\end{table}

\section{\label{conclusion} Conclusion}

We demonstrated how third-order statistics, in the form of the bicoherence and biphase, are useful in revealing noise transients with instances of nonlinear couplings among  various components of a detector. While for some transients the offending contributors to such couplings were identified here, nevertheless more work will be needed in the future to improve the fraction of successful identifications and establish terrestrial causes. 

It is important to stress that introducing artificial phase randomization in analyzing the data does not affect the actual phase coupling in the signal, but reduces the false detection of such couplings. We demonstrated this with simulations as well as an example of a glitch that appeared in real GW data. We used the PR auto-bicoherence to examine the similarities and differences in the phase-coupled frequencies within certain categories of glitches as well as across them. This third-order statistic provides an additional way of characterizing glitches than Omicron spectrograms and may contribute to troubleshooting their origins and subsequent removal or mitigation.

In future, we plan to study the cross-bicoherences of multiple auxiliary channels at the time of glitches. This will aid in identifying if nonlinear coupling of any of those channels are involved in producing the glitch. While conceptually its computation is not very different from that of auto-bicoherence, nevertheless sieving through dozens of auxiliary channel pairs significantly increases the computation load. New algorithms for computing it are being explored to find ways of reducing this load.

In addition to checking if certain noise transients, e.g., flagged by Omicron~\cite{Robinet-2020} or GravitySpy~\cite{Zevin-2017}, have origins in nonlinear coupling, bicoherence calculations can complement machine learning studies like DeepClean and NonSENS by confirming the presence of phase coupling as well as provide more information
about the nature of the nonlinear coupling (such as the specific frequencies and channels that
are coupled), which can be useful in regressing that excess noise in software and, in some cases, through hardware fixes. Conversely, phase couplings found first by bicoherence may be used for training machine learning algorithms to aid them in regressing excess GW strain noise arising from them. Whether other noise modeling and regression tools such as 
BayesWave~\cite{cornish:2014kda} and SHAPES~\cite{mohanty:2023mjn} can benefit from bicoherence computations remains to be seen. We will report on these investigations in a future work~\cite{bricewilliams}.

\section*{\label{Software details}{A note on software used and developed}}

We developed a customized Python \cite{Python} based Bicoherence (PyBicoh) analysis package for deployment in GW strain data~\cite{pybicoh}. It was used for producing the various auto-bicoherence maps and results discussed in this paper. We used GwPy~\cite{gwpy} to fetch and read GW data. PyCBC~\cite{alexnitz} functions were used to simulate CBC waveforms
and detector noise time-series (from a specificed noise PSD). All the figures in this paper are plotted with matplotlib~\cite{Hunter-2007}, with support from scipy~\cite{2020SciPy-NMeth}, numpy~\cite{harris2020array}, astropy~\cite{astropy:2013}, and h5py~\cite{andrewcollette}.


\section*{\label{acknowledgements} Acknowledgements}
We thank Andrew Lundgren, Fred Raab, Jess Mclver, Robert Bruntz, T. R. Saravanan, Shivaraj Kandhasamy, Suresh Doravari, Sanjit Mitra, Siddharth Soni, and Sunil Choudhary for discussions. Thanks are due to Miriam Cabero for  discussions on blip glitches. Deepak Bankar and Malathi Deenadayalan provided valuable computing-cluster support. We thank Ronaldas Macas for carefully reading the manuscript and sharing his comments on it. SS would like to thank Michał Bejger and acknowledge support provided by Tata Trusts and LIGO-India grants at IUCAA and  NCN-OPUS grant 2021/43/B/ST9/01714. We  are also grateful for the support provided by the Sarathi computing cluster (IUCAA) and the computational resources provided by the LIGO Laboratory supported by National Science Foundation Grants PHY-0757058 and PHY-0823459. This material is based upon work supported by NSF's LIGO Laboratory, a major facility fully funded by the National Science Foundation. We also acknowledge support from the NSF under grant PHY-2309352.

\begin{appendix}
\appendix

\section{\label{app:App-D} The bicoherence plane}
\begin{figure}[htp!]
	\centering
		\includegraphics[scale=.5]{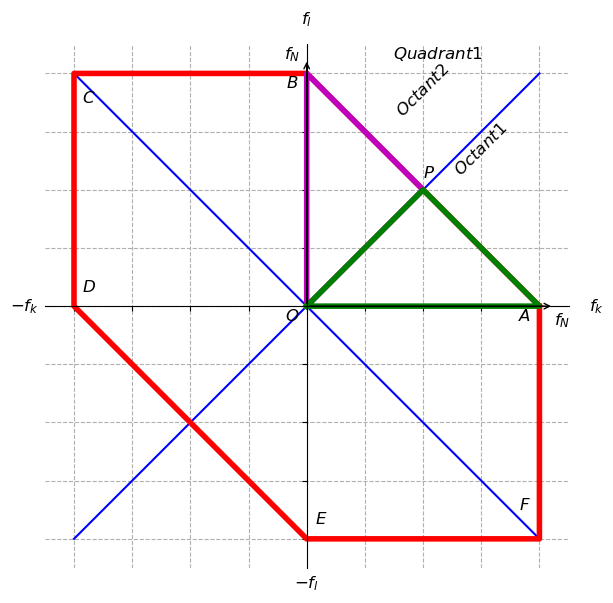}
		\caption{ The hexagon $ABCDEF$ in the bicoherence plane. The regions $\bigtriangleup OPB $ (magenta) and $ \bigtriangleup OPA $ (green) in Quadrant 1 are confined within the Nyquist frequency ($f_N$). }
		\label{fig:bicoherence_plane}
\end{figure}

In Fig.~\ref{fig:bicoherence_plane}, the part of the plane enclosed by the hexagon $ABCDEF$ is a valid bicoherence analysis section. Sections of quadrant 1 of this plane are considered for the bicoherence analysis in this work and displayed in our bicoherence maps. The region $\bigtriangleup OPA$ (green borders) is the principal region of non-redundant bispectral information. Other regions, such as $ \bigtriangleup OPB$ (magenta borders), have the same information. Specifically, patterns in  $ \bigtriangleup OPB$ of a bicoherence map will be identical to those in $\bigtriangleup OPA$ but with  reflection on the line $OP$. Note that the frequencies in these regions are confined within the Nyquist frequency ($f_N$). 

\section{\label{app:App-E} Similar bicoherence morphology}
\begin{figure}[htp!]
	\centering
		\includegraphics[scale=.7]{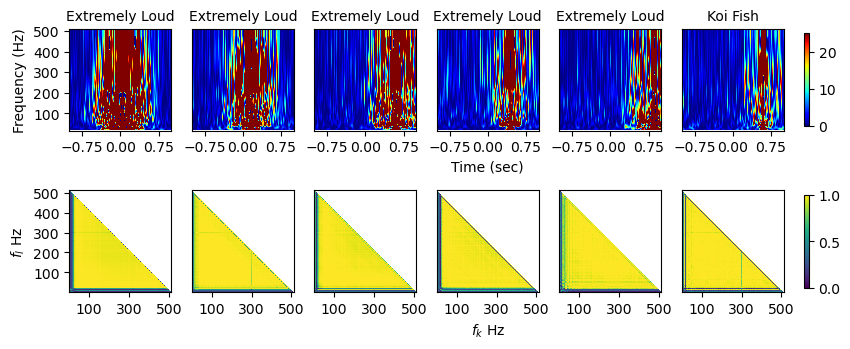}
        \caption{Top row: Spectrograms of ``extremely loud'' glitches (columns 1-5) and koi fish (column 6). Bottom row: Phase randomized bicoherence maps of corresponding glitches with 4sec data duration. Bicoherence maps in each sub-category of glitches reveal same phase-coupled bifrequencies, i.e., same bicoherence patterns, and suggest that the sources of these glitches are coupled in similar ways and manifested in the strain data.}
        \label{fig:3rdorder-2}
\end{figure}

A few examples of glitches with similar morphology in the third-order statistic are shown in Fig.~\ref{fig:3rdorder-2}. The top row there displays the time-frequency maps of those glitches and is classified by Gravity Spy~\cite{Zevin-2017} into the ``extremely loud'' and ``koi fish'' categories. The bottom row shows the corresponding bicoherence maps. 

\section{\label{app:C1} Bicoherence maps of some short duration compact binary coalescence signals}

\begin{figure}[htp!]
	\centering
		\includegraphics[scale=.85]{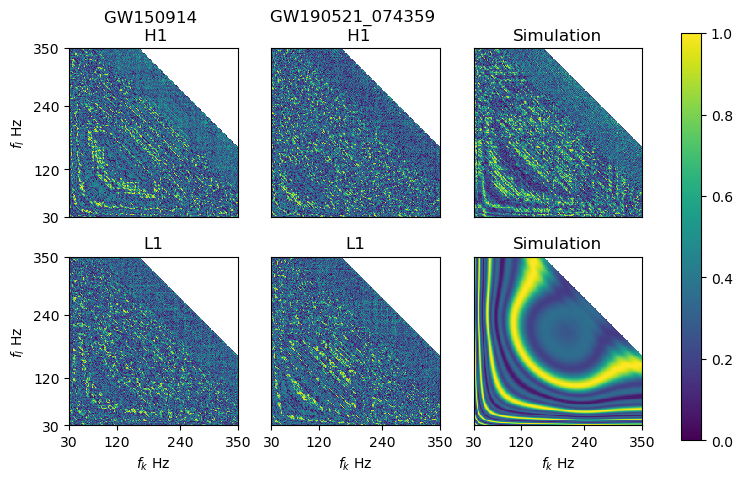}
		\caption{Phase-randomized bicoherence maps of the data from the binary black hole (BBH) signals GW150914 (column 1) and GW190521\_074359 (column 2), as well as a simulated BBH signal (column 3). 
		The bicoherence maps in the left and middle columns show interesting yet expected similarities between H1 and L1 signals. The top right figure shows such a map for a simulated GW150914-like signal added to detector noise, which itself is simulated using the noise power-spectral density of H1 data from around that event. The bottom right figure shows the same simulated GW150914-like signal but with no  noise added. Note that the bicoherence patterns here are in themselves not evidence for the presence of phase coupling between detector systems at that time. Rather, these patterns are characteristic of binary waveforms.  In identifying glitches by using their {\em auto}-bicoherence maps, it will be important to utilize this knowledge so as not to confuse real CBC signals for glitches.}
		\label{fig:gwevents}
\end{figure}

By applying the PR auto$-$bicoherence analysis to GWTC-3 \cite{GWTC-3} catalog events
we examine the apparent phase coherence in the data with CBC signals. We noticed 
significant bicoherence points in a few GW events. Two of those, namely, GW events GW150914 \cite{gw150914}, and  GW190521\_074359 \cite{GWTC-2} are presented in Fig.~\ref{fig:gwevents} 
along with a simulated signal for aiding the interpretation of their bicoherence maps. The presence of significant bicoherence is evident there. To prepare the GW strain data for the bicoherence analysis, first a high-pass filter is applied with cut-off frequency 20~Hz, followed by whitening and the application of a bandpass filter, from 30 to 350~Hz, with notches of the first three power line harmonics~\cite{Nielsen-2019}. The noise power-spectral density (PSD) is estimated from 32~sec of data  using the Welch method. 

To ensure that the phase couplings revealed in GW events' bicoherence maps are not from glitches, we modeled two signals similar to the GW150914 event, with and without strain noise. In Fig.~\ref{fig:gwevents} (row 1, column 3), the 8~second long time-series noise was 
simulated by using the H1 noise PSD available in Ref.~\cite{psd} for the GW190514 event. 
A simulated CBC signal was added to that noise before producing the plot. On the other hand, in Fig.~\ref{fig:gwevents} (row 2, column 3) no noise was added to the CBC signal. The bicoherence patterns in the map of the modeled signal, with added simulated noise, resembles those of the GW150914 signal, as shown in Fig.~\ref{fig:gwevents} (row 1, column 3). 
These signals were produced from the SEOBNRv4\_opt waveform~\cite{Bohe:2016gbl}, with component masses of 36 M\textsubscript{\(\odot\)} and 30 M\textsubscript{\(\odot\)}. This family of waveforms was constructed in the literature by calibrating effective one-body waveforms with an extensive set of numerical relativity waveforms, and have been employed in LIGO-Virgo CBC searches (see, e.g., Ref.~\cite{LIGOScientific:2018mvr}). The matched-filter SNR of the signal is 20. At the very least, the knowledge of these bicoherence patterns will be important in not confusing real CBC signals for noise artifacts and, therefore, for the purpose of veto safety. We exphasize that bicoherence maps are not the optimal method for detecting such CBC signals; rather the matched-filtering technique is~\cite{creighton}.

\section{\label{app:App-F} Cross-bicoherence: A feasibility study}

As noted above, the cross-bicoherence can help in finding the cause of glitches that arise due to the nonlinear coupling of disturbances of a pair of detector systems or components. However, cross-bicoherence is expensive to compute because it requires using pairs selected from hundreds of channels when little is known about the origins of a glitch. On the other hand, when some prior information is available on the origins, then the computational cost can be reasonable. For example, fast and slow scattered-light glitches and their origins are reasonably well understood~\cite{Soni_2021,Soni_2023,Tolley_2023}. Here, we pursue a feasibility study of cross-bicoherence with a small sample of slow scatter  glitches. 

One of the reasons for our choosing to study slow scatter glitches is that they are short -- of a median duration $\sim 3.2$-sec, which is of a timescale similar to the window we have tuned our bicoherence analysis on. The other reason is that their cause is now better understood: Low-frequency ground-motion, in the 0.03-0.1~Hz band, or micoseismic noise, in the 0.1-0.3~Hz band (for slow scatter glitches) or anthropogenic noise in 1-6~Hz (fast scatter glitches) couple with high quality-factor resonances of the detector to create upconverted noise in the 10-100~Hz detector band that sometimes rises to frequencies as high as $\sim 400$~Hz~\cite{Soni_2023}. Baffles that are installed in multiple places in the detectors to prevent scattered light from combining back with the main beam can actually play truant when not properly damped, and can amplify incident disturbances at their mechanical resonances. In particular, the cryo-manifold baffles in Hanford and the arm cavity baffles in Livingston have been shown to be involved in causing the fast scatter glitches at those respective sites. On the other hand, light scattering from the annular end-reaction mass (AERM) and the transmission monitor system (TMS) have been implicated as the cause of slow scatter glitches. 

\begin{figure}[htp!]
	\centering
		\includegraphics[scale=.7]{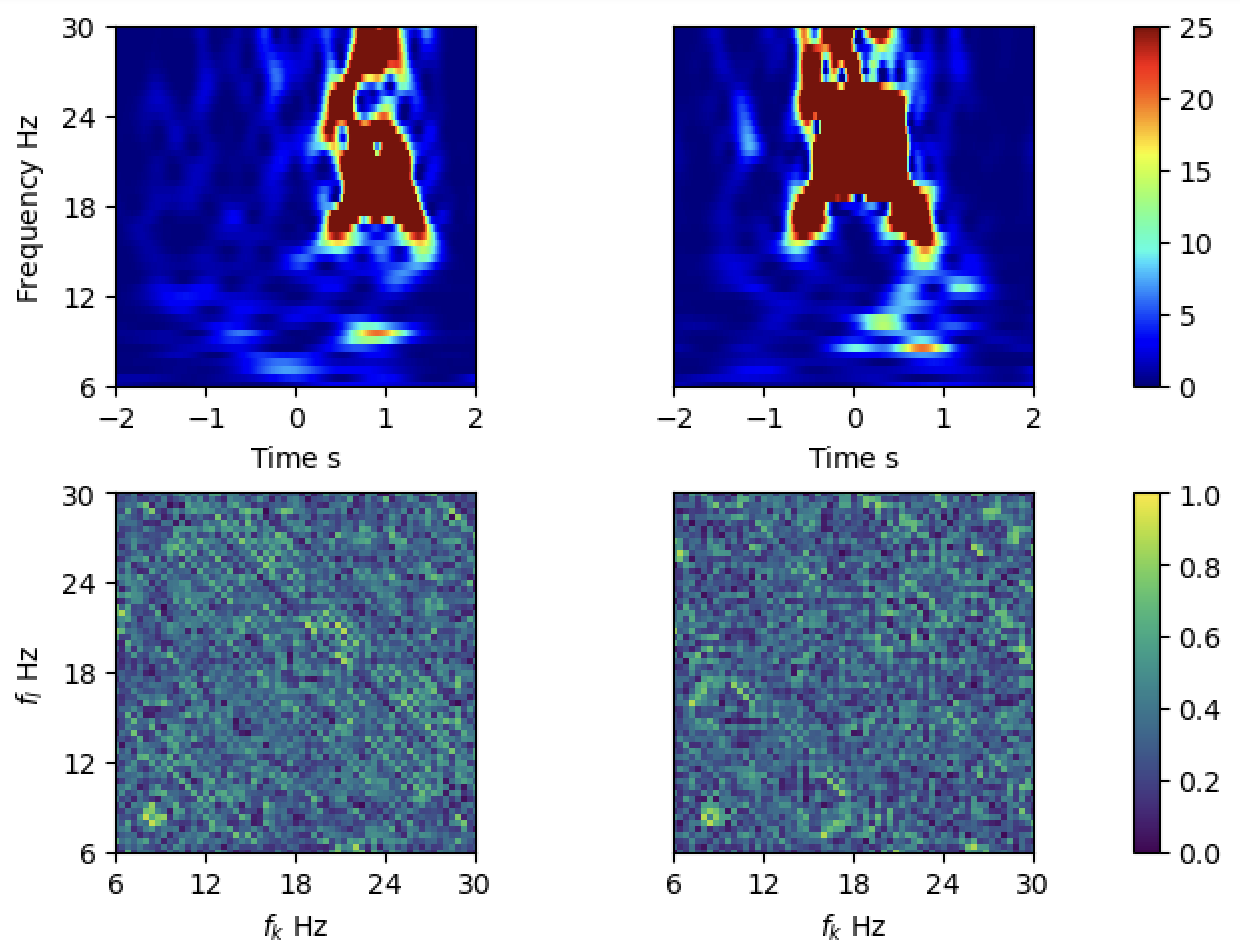}
		\caption{The spectrograms (top panel) and bicoherence maps (bottom panel) of two slow scattering glitches in LIGO-Livingston during O3. The spectrograms show the broadband power above $\sim16$~Hz over a $\sim2$~sec duration, as is characteristic of such glitches. The excess power at $\sim10$~Hz is known to be due to scattered-light associated with ETM-TMS noise coupling~\cite{Soni_2021}. The excess bicoherence at that frequency in the lower plots suggests the presence of nonlinear coupling and will be investigated in future.
}	\label{fig:fast_scattering_tf_cross}
\end{figure}

In Fig.~\ref{fig:fast_scattering_tf_cross} we show the spectrograms (top panel) and bicoherence maps (bottom panel) of two of the slow scattering glitches in LIGO-Livingston during O3 that we studied. The spectrograms show the broadband power above  $\sim16$~Hz over a
$\sim2$~sec duration, as is characteristic of such glitches. However, they also show excess noise around 10~Hz, which is known to be due to scattered-light associated with noise coupling between end test mass (ETM) and TMS~\cite{Soni_2021}. To compute the cross-bicoherences at those times, we take the $x(t)$ and $y(t)$ in Eq.~(\ref{eq:quadratic_equation}) to be the L1:ASC-X\_TR\_A\_NSUM\_OUT\_D channel, which is the most  relevant scattered light witness channel that is also publicly available. We took $z(t)$ to be the strain channel. We compute the cross-bicoherence with the same input parameters as mentioned in Sec.~\ref{sec:hos_qpc}. Interestingly, the neighborhood of that frequency in the bicoherence plots, i.e., $f_{k,l} \approx 10$~Hz, also shows significant bicoherence, which suggests the presence of nonlinear coupling. In a future work, we will report on a more elaborate investigation of noise couplings involved in scattered-light glitches,  including the 10~Hz feature. The quest for improving low-frequency sensitivity to GW signals may benefit from the modeling and removal of such noise as well.

\end{appendix}

\clearpage
\section*{References}
\bibliographystyle{iopart-num}
\bibliography{bicoherence}

\begin{thebibliography}{10}
\bibitem{pybicoh}
Pybicoh: A python based bicoherence analysis package for gravitational wave
  data: Package is yet to be released
\end{thebibliography}
\newpage

\end{document}